\documentclass[fleqn,10pt]{wlscirep}
\usepackage{caption}
\usepackage{subcaption}
\usepackage{multirow}
\usepackage[titletoc,title]{appendix}
\usepackage{tabularx}
\usepackage{blindtext} 

\title{Machine learning approach for early detection of autism by combining questionnaire and home video screening}

\author[1]{Halim Abbas}
\author[2]{Ford Garberson}
\author[3]{Eric Glover}
\author[4]{Dennis P Wall}
\affil[1,2]{Cognoa Inc., Palo Alto, CA, USA}
\affil[1]{www.linkedin.com/in/halimabbas}
\affil[3]{eric\_g@ericglover.com}
\affil[4]{Stanford University, Stanford, CA, USA}

\begin{abstract}
Existing screening tools for early detection of autism are expensive, cumbersome, time-intensive, and sometimes fall short in predictive value. In this work, we apply Machine Learning (ML) to gold standard clinical data obtained across thousands of children at risk for autism spectrum disorders to create a low-cost, quick, and easy to apply autism screening tool that performs as well or better than most widely used standardized instruments. This new tool combines two screening methods into a single assessment, one based on short, structured parent-report questionnaires and the other on tagging key behaviors from short, semi-structured home videos of children. To overcome the scarcity, sparsity, and imbalance of training data, we apply creative feature selection, feature engineering, and novel feature encoding techniques. We allow for inconclusive determination where appropriate in order to boost screening accuracy when conclusive. We demonstrate a significant accuracy improvement over standard screening tools in a clinical study sample of 162 children. 
\end{abstract}

\begin{document}

\flushbottom
\maketitle
\thispagestyle{empty}

\section{Introduction} \label{sec:Introduction}

Patient data pose unique challenges for Machine Learning. Of particular note is the application of machine learning to designing simple, low-cost screening tools for a variety of medical conditions. Labeled data usually originates from tightly controlled clinical environments and is hence clean but sparse, unbalanced, and of a different context to the data available when applying the screening techniques. This limits the reliability and generalizability of ML models trained on such data when applied in the less controlled settings of a typical screening tool. In this paper we highlight these challenges and present useful strategies for addressing them, focusing on the critical problem of affordable, reliable, early detection of autism in children 2-6 years old. 

Diagnosis within the first few years of life dramatically improves the outlook of children with autism, as it allows for treatment while the child's brain is still rapidly developing~\cite{bib:DurkinCrossSecStudy,bib:CDCAutismRates}. Unfortunately, in the United States autism is typically not diagnosed earlier than age 4, with approximately 27\% of cases remaining undiagnosed at age 8~\cite{bib:PediatricsAutismToddlerManagement}. This delay in diagnosis is driven primarily by a lack of effective screening tools and a shortage of specialists to evaluate at risk children.

Most autism screeners in use today are based on score-sheets with questions for the parent or the medical practitioner, that produce results by comparing summation scores to predetermined thresholds. Notable examples are the Modified Checklist for Autism in Toddlers, Revised (M-CHAT)~\cite{bib:MCHAT}, a checklist-based screening tool for autism that is intended to be administered during developmental screenings with children between the ages of 16 and 30 months, and the Child Behavior Checklist (CBCL)~\cite{bib:CBCLYoung} which is a parent-completed screening tool. 

Cognoa~\cite{bib:Cognoa} provides a data driven alternative screening method powered by machine learning and validated by multiple clinical studies ~\cite{bib:Duda2016, bib:ClinicalStudy2016}. To date, Cognoa has been used by over 150,000 parents in the US and internationally using web and native smartphone mobile app tools. The majority of Cognoa users are parents of young children between 18 and 30 months.

In this paper we present two new machine learning screeners that are reliable, cost-effective, easy enough to complete in minutes, and achieve higher accuracy than existing screeners. One is based on a short questionnaire about the child which is answered by the parent. The other is based on identification of specific behaviors by trained analysts after watching two or three short videos of the child at their home environment captured by parents using a mobile device. 

The parent questionnaire screener keys on behavioral patterns typically probed in a standard autism diagnostic instrument, the Autism Diagnostic Interview – Revised (ADI-R)~\cite{bib:ADIR}. This clinical tool consists of a parent interview of 93 multi-part questions with multiple choice and numeric responses which are delivered by a trained professional in a clinical setting. While this instrument is considered a gold-standard, and gives consistent results across examiners, the cost and time to administer it can be prohibitive. Section~\ref{sec:Questionnaire} details our approach to using clinical ADI-R instrument data to create a screener based on a much shorter questionnaire presented directly to parents without supervision.

The video screener keys on behavioral patterns typically probed in another diagnostic tool, the Autism Diagnostic Observation Schedule (ADOS)~\cite{bib:ADOS}. ADOS is widely considered a gold standard and is one of the most common behavioral instruments used to aid in diagnosis of autism~\cite{bib:MultistateStudyAutism}. It consists of an interactive, highly standardized examination of the child by trained clinicians in a tightly controlled setting. The ADOS is a multi-modular diagnostic instrument, with different modules for subjects in different developmental stages. Section~\ref{sec:Video} details our approach to using ADOS clinical data to create a video-based screener that relies on analysts evaluating short videos of children filmed by their parents at home. 

The use of ADI-R and ADOS score-sheets as inputs to train autism screening classifiers was introduced, studied, and clinically validated in previous work~\cite{bib:OrigQuestionnaire, bib:OrigVideo, bib:QuestionnaireValidation, bib:VideoStudies}. This paper builds upon that work, devising new algorithms that are more accurate and more robust against confounding biases between training and application data, and describing the improvements in the context of best practices for machine learning as applied to clinical science.

In Section ~\ref{sec:Combination}, we combine the parent questionnaire and video-based screeners into a single, higher reliability screening tool. Techniques for adapting the binary screeners to allow for a third, ``inconclusive" determination are also presented, allowing higher screening accuracy for those children who do receive ``autism" or ``not-autism" diagnoses at the cost of lower effective coverage. 

Finally, the screeners are applied to a sample of 162 at-risk children who have undergone full clinical examination and received a clinical diagnosis. The performance of our new screening tools surpasses the more conventional MCHAT and CBCL, and the combination screener is shown to achieve higher accuracy than either individual screener.

\section{Data} \label{sec:Data}
Training data were compiled from multiple repositories of ADOS and ADI-R score-sheets of boys and girls at 18 to 84 months of age including Boston Autism Consortium, Autism Genetic Resource Exchange, Autism Treatment Network, Simons Simplex Collection, and Vanderbilt Medical Complex. Since such repositories are highly imbalanced towards the incidence of autism, the controls across the datasets were supplemented with balancing data obtained by conducting ADI-R interviews on a random sample of children deemed at low risk for autism from Cognoa's user base.

The clinical validation sample consisted of 230 children who presented to one of three autism centers in the United States at 18 to 72 months of age. Every child received an ADOS as well as standard screeners like M-CHAT and CBCL as appropriate, and a diagnosis was ultimately ascertained by a licensed psychologist. This sample corresponds to a clinical study performed by Kanne et. al.~\cite{bib:ClinicalStudy2016} in 2016. For 162 of those children, the parents also used their mobile devices to complete the short parental questionnaire and submit the short videos required for the screeners discussed in this paper. 
Sample breakdown by age group and diagnosis for both the training and clinical validation datasets is shown in Table~\ref{tab:DataCharacteristics}.\textbf{}

\begin{table}[ht] 
\footnotesize
\centering
\begin{tabular}{|c|ccc|ccc|ccc|}
\hline
 & \multicolumn{3}{c|}{Parental questionnaire training} & \multicolumn{3}{c|}{Video analyst questionnaire training} & \multicolumn{3}{c|}{Clinical validation} \\
Age group & autism & other condition & neurotypical & autism & other condition & neurotypical & autism & other condition & neurotypical \\
\hline
Classification label &          + &                   - &                - &          + &                   - &                - &          + &                   - &                - \\
$< 4$ yr           &        414 &                 133 &               74 &       1445 &                 231 &              308 &         84 &                  18 &                3 \\
$\geq 4$ yr            &       1885 &                 154 &               26 &       1865 &                 133 &              277 &         37 &                  11 &                9 \\
\hline
\end{tabular}

\caption{\label{tab:DataCharacteristics} Dataset breakdown by age group and condition type for each of the sources of training data and for the clinical validation sample. The negative class label includes normally developing (i.e. neurotypical) children as well as children with  developmental delays and conditions other than autism.}
\end{table}
\normalsize

\section{Approach} \label{sec:Approach}
We trained two independent ML classifiers and combined their outputs into a single screening assessment. The parent questionnaire classifier was trained using data from historical item-level ADI-R score-sheets with labels corresponding to established clinical diagnoses. The video classifier was trained using ADOS instrument score-sheets and diagnostic labels. In each case, progressive sampling was used to verify sufficient training volume as detailed in Appendix \ref{sec:ProgressiveSampling}.

The ADI-R and ADOS instruments are designed to be administered by trained professionals in highly standardized clinical settings and typically take hours. Contrastingly, our screening methods are deliberately designed to be administered at home by parents without expert supervision, and to take only minutes to complete. This change of environment, further detailed in Appendix \ref{sec:ChangeOfEnvironment},  causes significant data degradation and biases resulting in an expected loss of screening accuracy. For each classifier, mindful adjustments to ML methodology were used to mitigate these issues, as explained below.

\subsection{Parent questionnaire}\label{sec:Questionnaire}

Multiple model variants representing incremental improvements over a generic ML classification approach are shown below. Each model was independently parameter-tuned with a bootstrapped grid search. Class labels were used to stratify the cross-validation folds, and (age, label) pairs were used to weight-balance the samples.  

\subsubsection{Generic ML baseline variant}\label{sec:NaiveQuestionnaireLearning}

A random forest was trained over the ADI-R instrument data. Each of the instrument's 155 data columns was treated as a categorical variable and one-hot encoded. The subject's age and gender were included as features as well. Of the resulting set of features, the top 20 were selected using feature-importance ranking in the decision forest.

\subsubsection{Robust feature selection variant}\label{sec:StableFeatureSelection}

Due to the small size and sparsity of the training dataset, generic feature selection over hundreds of binary features was not robust, and the selected features (along with the performance of the resulting model) fluctuated from run to run due to the stochastic nature of the learner's underlying bagging approach. Many ADI-R questions are highly correlated, leading to multiple competing sets of feature selection choices that were seemingly equally powerful during training, but which had different performance characteristics when the underlying sampling bias was exposed via full bootstrapped cross-validation. This resulted in a wide performance range of the variant in Section~\ref{sec:NaiveQuestionnaireLearning} as shown in Table \ref{tab:TrainingPerformances}.

Robust feature selection overcame that limitation using a two-step approach. First, a 100-count bootstrapped feature selection was run, with a weight balanced 90\% random sample selected in each iteration. The top 20 features were selected each time, and a rank-invariant tally was kept for the number of times each feature made it to a top-20 list. Next, the top 30 features in the tally were kept as candidates and all other features were discarded. A final feature-selection run was used to pick the best 20 of these candidate features. Over multiple runs, this approach was found to be more robust to statistical fluctuations, usually selecting the same set of features when run multiple times.

\subsubsection{Age silo variant}\label{sec:BinningByAge}

This variant built upon the improvements of Section~\ref{sec:StableFeatureSelection}, by exploiting the importance of the dichotomy between pre-phrasal and fully-phrasal language capability in at-risk children.  Language development is significant in this domain as it is known to affect the nature in which autism presents, and consequently the kinds of behavioral clues to look for in order to screen for it.

This variant achieved better performance by training separate classifiers for children in the younger and older age groups of Table~\ref{tab:DataCharacteristics}. The age dichotomy of $<4$, $\geq4$ was chosen to serve as the best proxy for language ability. Feature selection, model parameter-tuning, and cross-validation were run independently for each age group classifier. Before siloing by age group, the classifier was limited to selecting features that work well across children of both developmental stages. Siloing enabled the classifiers to specialize on features that are most developmentally appropriate within each age group.

\subsubsection{Severity-level feature encoding variant}\label{sec:Encoding}

Building upon Section~\ref{sec:BinningByAge}, this variant achieved better performance by replacing one-hot feature encoding with a more context-appropriate technique. One-hot encoding does not distinguish between values that correspond to increasing levels of severity of a behavioral symptom, and values that do not convey a clear concept of severity. This is especially troublesome since a typical ADI-R instrument question includes answer choices from both types of values. For example, ADI-R question 37, which focuses on the child's tendency to confuse and mix up pronouns, allows for answer codes 0,1,2,3,7,8, and 9. Among those choices, 0 through 3 denote increasing degrees of severity in pronominal confusion, while 7 denotes any other type of pronominal confusion not covered in 0-3 regardless of severity. Codes 8 and 9 denote the non-applicability of the question (for example to a child still incapable of phrasal speech) or the lack of an answer (for example if the question was skipped) respectively. When coding the answers to such question, generic one-hot encoding would allow for non-symptomatic answer codes to be selected as screening features based on phantom correlations present in the dataset. 

Severity-level encoding converts all answer codes that do not convey a relevant semantic concept to a common value, thereby reducing the chance of useless feature selection, and reducing the number of features to choose from. In addition, severity-level encoding condenses the signal according to increasing ranges of severity. This more closely resembles the way medical practitioners interpret such answer choices in real life, and helps alleviate the problem of sparsity over each of the one-hot encoded features in the dataset. Severity-grouped feature levels also reduce over-fitting and serve to emphasize the patterns of correlation between the features and the target, with signals from multiple answer codes grouped into single binary features in a semantically-appropriate way. The associated severity-level encoding for the possible responses to the exemplary ADI-R question 37 is shown in Table~\ref{tab:ADIExampleEncoding}.

\begin{table}[!h]
\centering
\begin{tabular}{l|c|c|c|c|c|c|c}
\hline
Original answer & 0 & 1 & 2 & 3 & 7 & 8 & 9 \\
Severity encoded features & $=0$ & $\ge1$ & $\ge1$ and $\ge2$ & $\ge1$ and $\ge2$ and $\ge3$ & $=7$ & None & None \\
\hline
\end{tabular}
\caption{Illustration of the severity-level encoding technique on responses to ADI-R question 37. For example, a response of ``2" will set both the features of $\ge1$ and $\ge2$ to ``True''.}
\label{tab:ADIExampleEncoding}
\end{table}

\subsubsection{Aggregate features variant} \label{sec:FeatureEngineering}

Building upon Section~\ref{sec:Encoding}, this variant achieved better performance by incorporating aggregate features such as the minimum, maximum, and average severity level, as well as number of answer choices by severity level across the questions corresponding to the 20 selected features. These new features were especially helpful due to the sparse, shallow, and wide nature of the training set, whereupon any semantically meaningful condensation of the signal can be useful to the trained classifier. The summation of answer choices by severity is a method used by specialists who administer the ADI-R instrument in real life, and is therefore known to carry semantic benefit. 

\subsubsection{Inconclusive results variant}\label{sec:Inconclusive}

Patients with more complex symptom presentation are known to pose challenges to developmental screening. These children often screen as false positives or false negatives, resulting in an overall degradation of screening accuracy that is observed by all standard methods and has become acceptable in the industry. Given that our low-cost instruments do not rely on sophisticated observations to differentiate complex symptom cases, our approach was to avoid assessing them altogether, and to try instead to spot and label them as ``inconclusive".

Building upon \ref{sec:FeatureEngineering}, two methods to implement this strategy were devised. The first was to train a binary classifier with a continuous output score, then replace the cutoff threshold with a cutoff range, with values within the cutoff range considered inconclusive. A grid search was used to determine the optimal cutoff range representing a tradeoff between inconclusive determination rate and accuracy over conclusive subjects. The second approach was to train and cross-validate a simple binary classifier, label the correctly and incorrectly predicted samples as conclusive or inconclusive respectively, and then build a second classifier to predict whether a subject would be incorrectly classified by the first classifier. At runtime, the second classifier was used to spot and label inconclusives. The conclusives were sent for classification by a third, binary classifier trained over the conclusive samples only. Both methods for labeling inconclusive results yielded similar performance. Therefore the simpler method of using a threshold range in the machine learning output was used to report inconclusive results for this paper.

\subsection{Video} \label{sec:Video}

The second of our two-method approach to autism screening is an ML classifier that uses input answers about the presence and severity of target behaviors among subjects. Analysts provide this information upon viewing two or three 1-minute home videos of children in semi-structured settings, that are taken by parents on their mobile phones. The classifier was trained on item-level data from two of the ADOS modules (module 1: preverbal, module 2: phrased speech) and corresponding clinical diagnosis.

Two decision forest ML classifiers were trained corresponding to each ADOS module. For each classifier, 10 questions were selected using robust feature selection as in Section ~\ref{sec:StableFeatureSelection}, and allowance was made for inconclusive outcomes as in Section ~\ref{sec:Inconclusive}. Each model was independently parameter-tuned with a bootstrapped grid search. Class labels were used to stratify the cross-validation folds, and (age,label) pairs were used to weight-balance the samples.

Problems related to the change of environment from training to application (detailed in Appendix~\ref{sec:ChangeOfEnvironment}) are especially significant in the case of video screening because ADOS involves a 45 minute direct observation of the child by experts, whereas our screening was based on unsupervised short home videos. Specifically, we expect the likelihood of inconclusive or unobserved behaviors and symptoms to be much higher in the application than in the training data, and the assessed level of severity or frequency of observed symptoms to be less reliable in the application than in the training data. The following improvements were designed to help overcome these limitations.

\subsubsection{Presence of behavior encoding}\label{sec:PresenceEncoding}
To minimize potential bias from a video analyst misreading the severity of a symptom in a short cell phone video, this encoding scheme improves feature reliability at the expense of feature information content by collapsing all severity gradations of a question into one binary value representing the presence vs absence of the behavior or symptom in question. Importantly, a value of 1 denotes the presence of behavior, regardless of whether the behavior is indicative of autism or of normalcy. This rule ensures that a value of 1 corresponds to a reliable observation, whereas a 0 does not necessarily indicate the absence of a symptom but possibly the failure to observe the symptom within the short window of observation.

\subsubsection{Missing value injection to balance the non presence of features for the video screener training data}\label{sec:MissingValueInjectionWhenNotPresent}

While collapsing severity gradations into a single category overcomes noisy severity assessment, it does not help with the problem of a symptom not present or unnoticeable in a short home video. For this reason, it is important that the learning algorithm treat a value of 1 as semantically meaningful, and a value of 0 as inconsequential. To help accomplish that we augmented the training set with duplicate samples that had some feature values flipped from 1 to 0. The injection of 0s was randomly performed with probabilities such that the sample-weighted ratio of positive to negative samples for which the value of any particular feature is 0 about 50\%. Such ratios ensure that the trees in a random forest will be much less likely to draw conclusions from the absence of a feature.

\subsection{Combination} \label{sec:Combination}

The numerical response of each of the parent questionnaire and video classifiers were combined using logistic regression. Since each of the individual methods was siloed, separate combinators were trained per age group silo. For each combinator, optimal inconclusive output criteria were chosen as discussed Section~\ref{sec:Inconclusive}. The performance characteristics of the overall screening process compared to standard alternative screeners is shown in Section~\ref{sec:CombinationResults}.

\section{Results} \label{sec:Results}

\subsection{Parent questionnaire performance on training data}\label{sec:QuestionnairePerformance}

Bootstrapped cross-validation performance metrics for the optimally parameter-tuned version of each of the variants in~\ref{sec:Questionnaire} are reported in Table~\ref{tab:TrainingPerformances}. The results for baseline variant~\ref{sec:NaiveQuestionnaireLearning} are reported as a confidence interval rather than a single value, because the unreliability of generic feature selection leads to different sets of features selected from run to run, with varying performance results.

\begin{table}[]
\small
\centering
\begin{tabular}{p{3.5cm}|p{1cm}p{1cm}p{1cm}|p{1cm}p{1cm}p{1cm}|p{1cm}p{1cm}p{1cm}}
\hline
 & \multicolumn{3}{|c|}{AUC} & \multicolumn{3}{c|}{Sensitivity} & \multicolumn{3}{c}{Specificity} \\
                         scenario &        All&       $< 4$ yr&    $\geq 4$ yr& All& $< 4$ yr& $\geq 4$ yr&  All& $< 4 yr$& $\geq 4 yr$\\
\hline
                 Generic ML baseline &  0.932 to 0.950 &  0.928 to 0.953 &  0.928 to 0.953 &    0.976 to 0.982 &      0.975 to 0.984 &         0.975 to 0.984 &  0.628 to 0.645 &   0.625 to 0.648 &      0.625 to 0.648 \\
         Robust feature selection variant &          0.958 &           0.958 &           0.958 &             0.982 &               0.982 &                  0.982 &           0.624 &            0.624 &               0.624 \\
                         Age silo variant &          0.953 &           0.939 &           0.961 &             0.962 &               0.939 &                  0.977 &           0.777 &            0.774 &               0.779 \\
                    Severity-level feature encoding variant &          0.965 &            0.950 &           0.974 &              0.960 &               0.912 &                  0.993 &           0.748 &            0.833 &               0.692 \\
         Aggregate features variant &          0.972 &           0.987 &           0.963 &             0.992 &               0.988 &                  0.994 &           0.754 &            0.894 &               0.661 \\
 With inconclusive allowance [up to 25\%]  &          0.991 &           0.997 &           0.983 &                 1.000 &                   1.000 &                      1.000 &           0.939 &            0.977 &               0.881 \\
\hline
\end{tabular}

\caption{Cross validation performance of increasingly effective classifier variants based on the training data for the parent questionnaire}
\label{tab:TrainingPerformances}
\end{table}

\subsection{Parent questionnaire performance on clinical data}\label{sec:QuestionnaireClinicalData}

Parents of children included in the clinical study answered short, age-appropriate questionnaires chosen using the feature selection approach outlined in~\ref{sec:StableFeatureSelection}. The performance metrics for each of the classification variants that build upon that feature selection scheme are shown in Table~\ref{tab:ClinicalPerformances}.

\begin{figure}[h]
\centering
 \includegraphics[width=\linewidth]{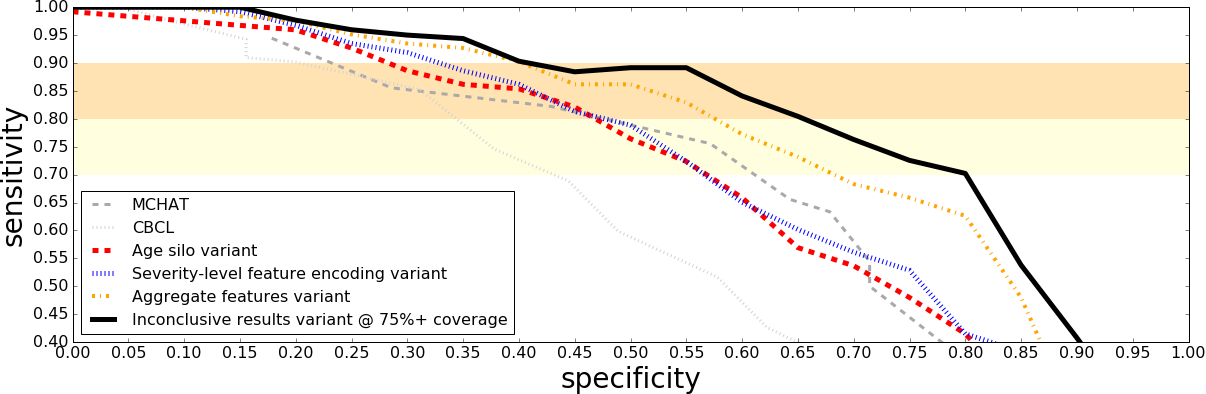}
 \caption{ROC curves on the clinical sample for various questionnaire based autism screening techniques. Note that unlike  Figures~\ref{fig:CombinedROCCurvesWithoutDunno} through~\ref{fig:CombinedROCCurvesWithDunno} and~\ref{fig:YoungCombinedROCCurvesWithoutDunno} through~\ref{fig:OldCombinedROCCurvesWithDunno}, 168 children are included in this sample (six children did not have videos available).}
\label{fig:QuestionnaireROCCurves}
\end{figure}

ROC curves in Figure~\ref{fig:QuestionnaireROCCurves} show how our parent questionnaire classification approach outperforms some of the established screening tools like MCHAT and CBCL on the clinical sample.

\begin{table}[]
\small
\centering
\begin{tabular}{p{3.5cm}|p{1cm}p{1cm}p{1cm}|p{1cm}p{1cm}p{1cm}|p{1cm}p{1cm}p{1cm}}
\hline
 & \multicolumn{3}{|c|}{AUC} & \multicolumn{3}{c|}{Average precision @ 80\% sensitivity} & \multicolumn{3}{c}{Specificity @ 80\% sensitivity} \\
scenario &  All &  $< 4$ yr&  $\geq 4$ yr&  All  &  $< 4$ yr &  $\geq 4$ yr &  All  &  $< 4$ yr  &  $\geq 4$ yr \\

\hline
Age silo variant                        &     0.62 &          0.68 &             0.54 &                           0.65 &                                0.62 &                                   0.52 &            0.48 &                 0.46 &                    0.24 \\
Severity-level feature encoding variant &     0.67 &          0.69 &             0.64 &                           0.64 &                                0.62 &                                   0.58 &            0.48 &                 0.46 &                    0.33 \\
Aggregate features variant              &     0.68 &          0.73 &             0.68 &                           0.68 &                                0.69 &                                   0.65 &            0.57 &                 0.62 &                    0.48 \\
With inconclusive allowance [up to 25\%]                     &     0.72 &          0.72 &             0.73 &                           0.70 &                                0.72 &                                   0.67 &            0.67 &                 0.71 &                    0.53 \\
\hline
\end{tabular}

3\caption{Performance over clinical sample of increasingly effective classifier variants based on the parent questionnaire, by age group. In all cases the metric threshold has been tuned to 80\% sensitivity.}
\label{tab:ClinicalPerformances}
\end{table}
\normalsize

\subsection{Combination screening performance on clinical data}
\label{sec:CombinationResults}

\begin{figure}[h]
\centering
 \includegraphics[width=\linewidth]{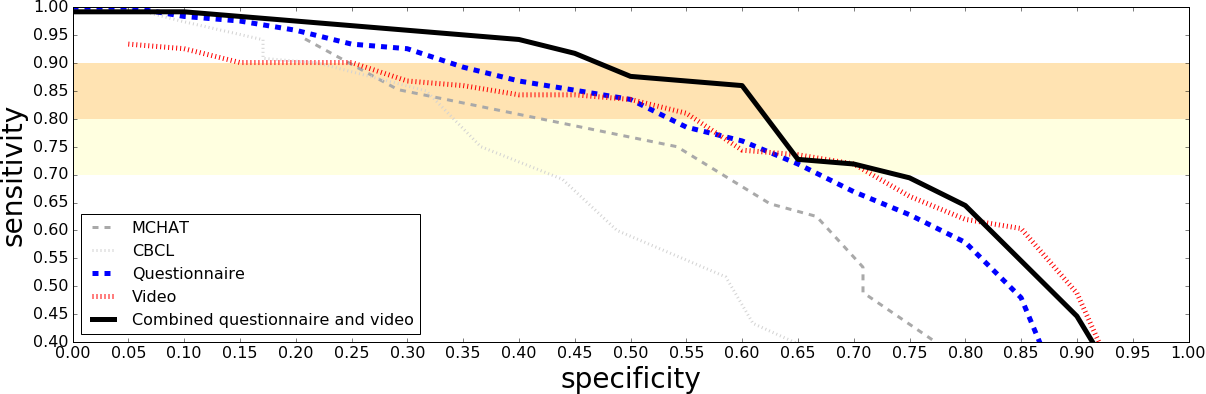}
 \caption{ROC curves on the clinical sample for the questionnaire and the video based algorithms, separately and in combination. The established screening tools MCHAT and CBCL are included as baselines.}
\label{fig:CombinedROCCurvesWithoutDunno}
\end{figure}

\begin{figure}[htbp]
\centering
 \includegraphics[width=\linewidth]{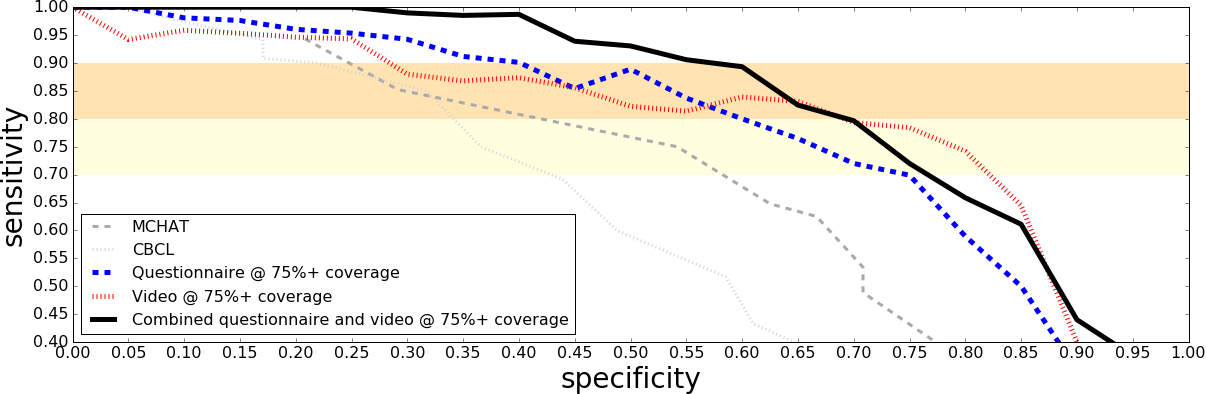}
 \caption{ROC curves on the clinical sample for the questionnaire and the video based algorithms, separately and in combination. Inconclusive determination is allowed for up to 25\% of the cases. The established screening tools MCHAT and CBCL are included as baselines.}
\label{fig:CombinedROCCurvesWithDunno}
\end{figure}

ROC curves in Figure~\ref{fig:CombinedROCCurvesWithoutDunno} show how combining the questionnaire and video classifiers into a single assessment further boosted performance on the clinical study sample. Figure ~\ref{fig:CombinedROCCurvesWithDunno} shows how allowing for inconclusive determination up to 25\% of the time results in further accuracy improvement over the conclusive cases.

\section{Conclusion} \label{sec:Conclusion}

Machine learning can play a very important role in improving the effectiveness of clinical screeners. We have achieved a significant improvement over established screening tools for autism in young children as demonstrated in a clinical trial. We have also shown some important pitfalls when applying machine learning in a clinical setting, and quantified the benefit of applying proper solutions to address them.

\vspace*{300pt}

\clearpage

\bibliography{references}

\clearpage

\begin{appendices}

\section{Progressive sampling on training datasets}\label{sec:ProgressiveSampling}

\begin{figure*}[ht]
    \centering
    \begin{subfigure}[t]{0.5\textwidth}
        \centering
        \includegraphics[width=\textwidth]{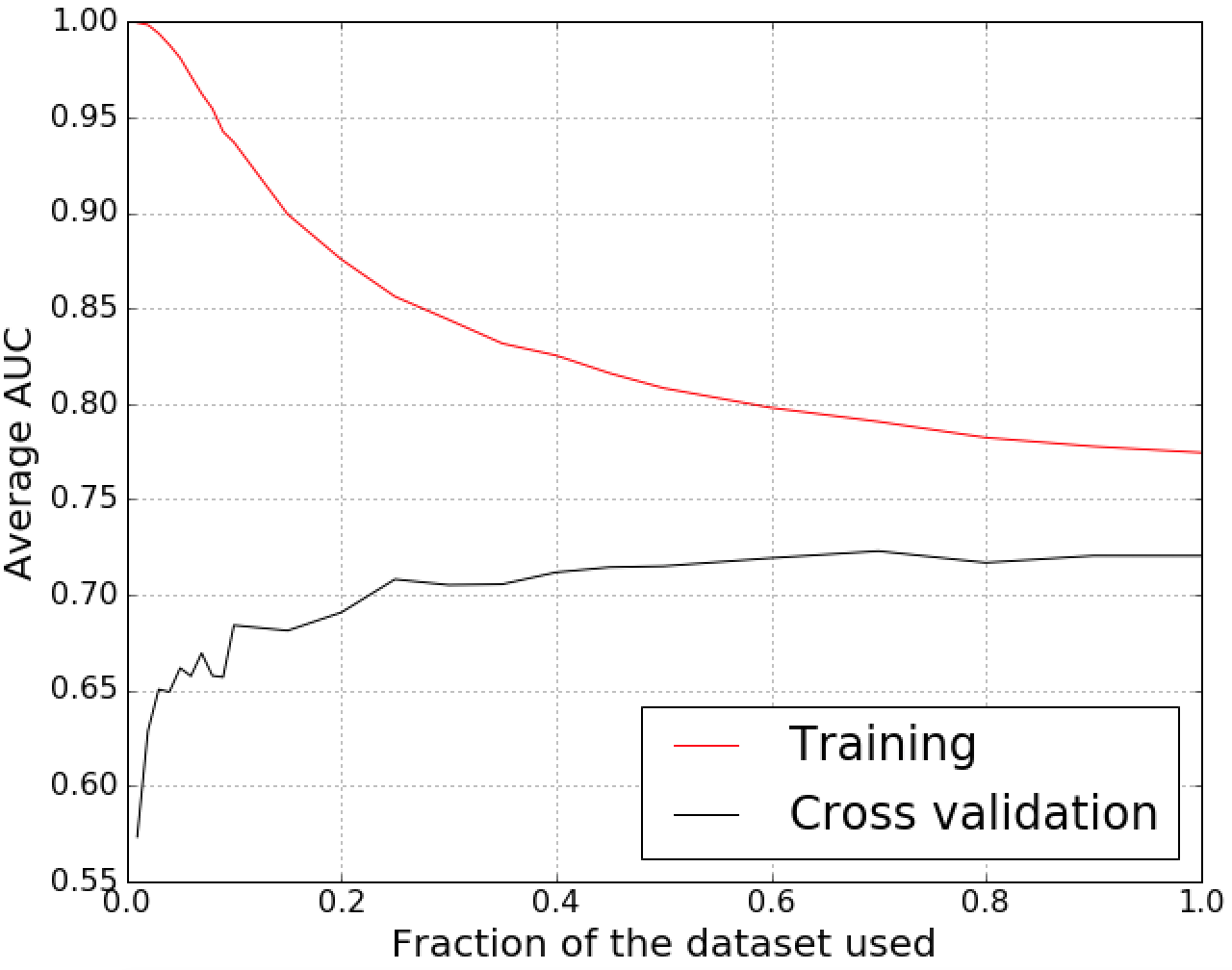}
        \caption{Subset of ADOS module 1 (generally pre-verbal) children.}
        \label{fig:StatLimitationsMod1}
    \end{subfigure}%
    ~
    \begin{subfigure}[t]{0.5\textwidth}
        \centering
        \includegraphics[width=\textwidth]{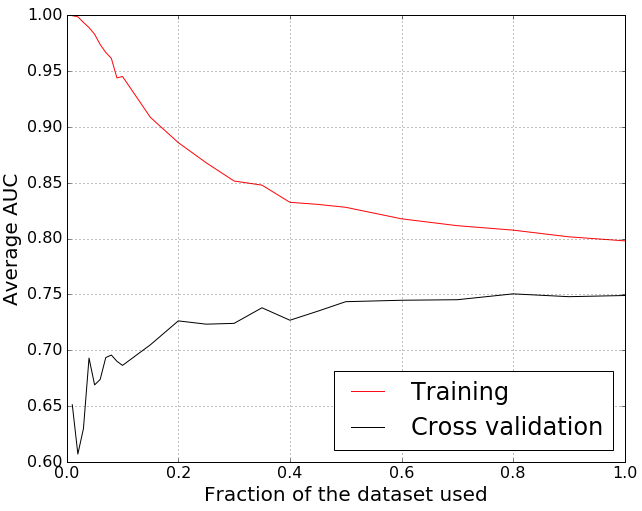}
        \caption{Subset of ADOS module 2 (generally verbal) children.}
        \label{fig:StatLimitationsMod2}
    \end{subfigure}
    \caption{Progressive sampling used to establish the availability of  sufficient training data for training an ML classifier using the ADOS-based training set.}
    \label{fig:StatLimitations}
\end{figure*}

Progressive sampling runs were performed to assert available training data is sufficient to build stable ML classifiers. These runs were performed for each classifier variant for both the questionnaire and video based training samples. In each run, boot-strapped cross-validation was used to compute the AUC metric of an optimized random-forest trained over increasingly larger proportions of the training set. The size of the training set was demonstrated to be sufficient for stable learning of ensemble decision trees, as shown in the plots of Figure~\ref{fig:StatLimitations}. A similar conclusion was reached for the parental questionnaire classifiers for young and old children, using similar progressive sampling runs of its corresponding training sets.

\section{Differences between training and application environments} \label{sec:ChangeOfEnvironment}

Since the classification labels used to train medical screening algorithms would have to be come from professional medical diagnoses, it is not feasible to amass large training sets that are class-balanced and sufficiently representative of the general population of children.
Our approach is to start with historical medical instrument records of previously diagnosed subjects, and use those as training data for screeners that will rely on information acquired outside the clinical setting. Since the algorithms are trained using features obtained during rigorous clinical evaluations, but applied on data representing proxies of said features acquired in a less controlled setting, we expect the performance of the classifiers to degrade considerably when used as a mobile screener.

More specifically, Table~\ref{tab:ApplicationSettingTable} details the various mechanisms by which confounding biases may creep into the application data in ways not present in the training data. It should be noted that inaccuracies introduced by such biases cannot be probed by cross validation or similar analysis of the training data alone.

\begin{table}[ht]
 \centering  
\begin{tabular}{p{5.4cm}|p{5.4cm}|p{5.4cm}}
\hline
 \multicolumn{1}{c|}{Aspect} & \multicolumn{1}{c|}{Training Setting} & \multicolumn{1}{c}{Application Setting} \\
\hline
Source &
 ADI-R and ADOS instrument administered by trained professionals during clinical evaluations &
 Short parent questionnaires displayed on smartphone, and behavior tagging by analysts after observing two or three 1-minute home videos uploaded by parents \\
 
 Proctor & 
 Highly trained medical professionals & 
 Parents answering the questionnaires are untrained, and the analysts evaluating the home videos are only minimally trained. As a result, their answers may not be as consistent, objective, or reliable \\

 Setting &
 Clinic setting with highly standardized and semi-structured interactions &
 At home. Not possible to recreate the structured clinical environment, resulting in an undesired variability of the output signals. Subjects might also behave differently at the clinic than at home, further amplifying the bias  \\

 Duration & 
 The ADI-R can take up to 4 hours to complete; The ADOS can take up to 45 minutes of direct observation by trained professionals & 
 Under 10 minutes to complete the parent questionnaire, and a few minutes of home video. As a result, some symptoms and behavioral patterns might be present but not observed. Also causes big uncertainty about the severity and frequency of observed symptoms  \\

 Questionnaires &
 Sophisticated language involving psychological concepts, terms, and subtleties unfamiliar to non experts  &
 Simplified questions and answer choices result in less nuanced, noisier inputs\\

\hline     
\end{tabular}
 \caption{\label{tab:ApplicationSettingTable} Differences between training and application environments. These differences are expected to cause bias that cannot be captured by cross validation studies.}
 \end{table}

\section{Age binned results}\label{sec:AppendixB}

\begin{figure}[ht]
\centering
 \includegraphics[width=\linewidth]{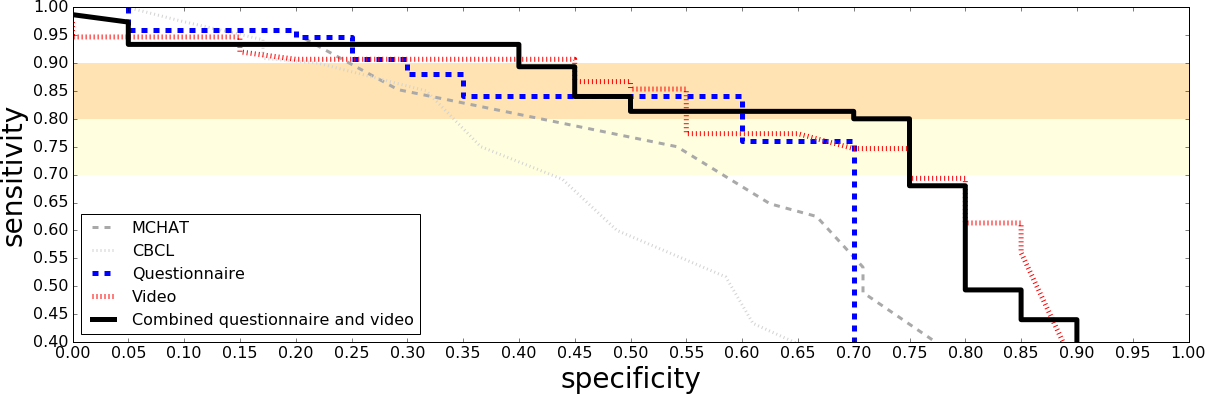}
 \caption{ROC curves on the clinical results for children under four years of age, for the questionnaire and the video based algorithms, as well as the combination. Comparisons with the the established (non machine learning) screening tools MCHAT and CBCL are also shown.}
\label{fig:YoungCombinedROCCurvesWithoutDunno}
\end{figure}

\begin{figure}[ht]
\centering
 \includegraphics[width=\linewidth]{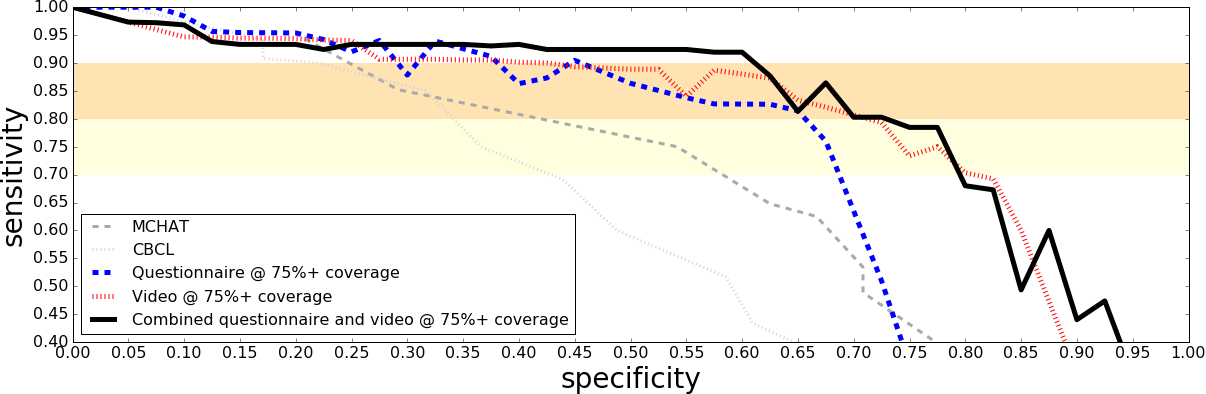}
 \caption{ROC curves on the clinical results for children under four years of age, for the questionnaire and the video based algorithms, as well as the combination, restricted to the children who were not determined to have an inconclusive outcome (tuned to have at most 30\% allowed to be inconclusive). Comparisons with the the established (non machine learning) screening tools MCHAT and CBCL are also shown.}
\label{fig:YoungCombinedROCCurvesWithDunno}
\end{figure}

\begin{figure}[ht]
\centering
 \includegraphics[width=\linewidth]{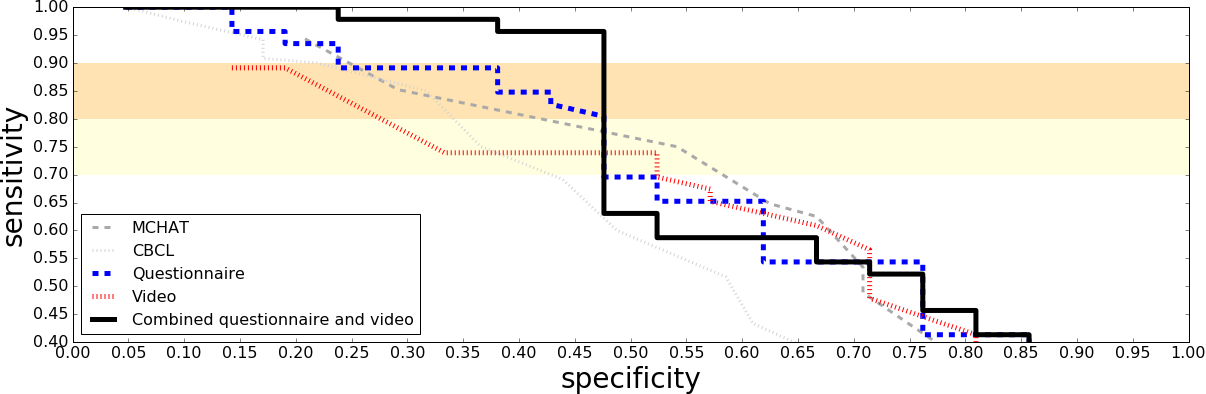}
 \caption{ROC curves on the clinical results for children of at least four years of age, for the questionnaire and the video based algorithms. Comparisons with the established (non machine learning) screening tools MCHAT and CBCL are also shown.}
\label{fig:OldCombinedROCCurvesWithoutDunno}
\end{figure}

\begin{figure}[ht]
\centering
 \includegraphics[width=\linewidth]{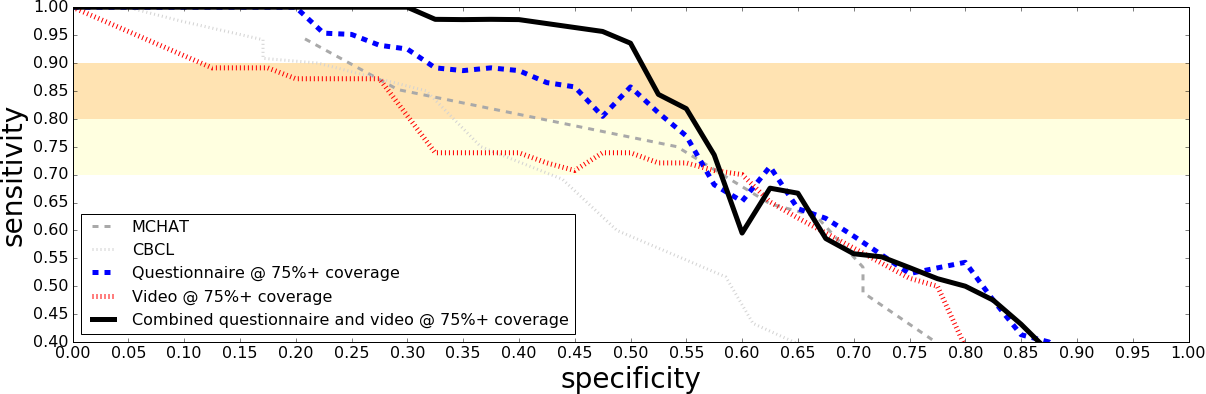}
 \caption{ROC curves on the clinical results for children of at least four years of age, for the questionnaire and the video based algorithms, as well as the combination, restricted to the children who were not determined to have an inconclusive outcome (tuned to have at most 30\% allowed to be inconclusive). Comparisons with the established (non machine learning) screening tools MCHAT and CBCL are also shown.}
\label{fig:OldCombinedROCCurvesWithDunno}
\end{figure}

Results split for young and old kids separately are shown in Figure~\ref{fig:YoungCombinedROCCurvesWithoutDunno} through ~\ref{fig:OldCombinedROCCurvesWithDunno}.



\clearpage

\end{appendices}

\end{document}